\documentclass[twocolumn,english]{article}
\usepackage[T1]{fontenc}
\usepackage[latin9]{inputenc}
\usepackage{verbatim}
\usepackage{amstext}
\usepackage{graphicx}
\usepackage{amsmath}

\makeatletter
\newcommand{\lyxaddress}[1]{
\par {\raggedright #1
\vspace{1.4em}
\noindent\par}
}

\usepackage{babel}

\newcommand{\ket}[1]{|#1 \rangle}

\usepackage{mathrsfs}

\makeatother

\usepackage{babel}
\begin{document}

\title{Error measurements for a quantum annealer using the one-dimensional Ising model with twisted boundaries}

\author{Nicholas Chancellor{*$^\%$}, Philip J. D. Crowley{*$^\pounds$}, Tanja \DJ{}uri\'{c}{*$^\$$},
Walter Vinci*$^\sim$\\
 Mohammad H. Amin''$^{\ddagger}$, Andrew G. Green{*}, Paul A. Warburton{*}\textasciicircum{},
and Gabriel Aeppli{$^\&$}}

\maketitle

\lyxaddress{{*}London Centre For Nanotechnology, University College London19 Gordon Street, London UK, WC1H
0AH\\
`` D-Wave Systems Inc. 3033 Beta Avenue Burnaby, British Columbia,
Canada V5G 4M9\\
{$^\&$}Physics Department, ETH Z{\"u}rich, Switzerland CH-8093; Institut de Physique, EPFL, Lausanne, Switzerland CH-1015; Paul Scherrer Institute, Villigen, Switzerland CH-5232; Quantum Center, Eidgen{\"o}ssische Technische Hochschule Zurich, CH-8093 Zurich, Switzerland\\
$^{\ddagger}$Department of Physics, Simon Fraser University, Burnaby,
BC Canada V5A 1S6\\
\textasciicircum{}Department of Electronic and Electrical Engineering,
UCL, Torrington Place London UK WC1E 7JE\\
$^\%$ current affiliation: Department of Physics, Durham Newcastle Joint Quantum Centre, Durham University, South Road, Durham, DH1 3LE\\
$^\pounds$current affiliation: Department of Physics, Boston University, Boston, MA 02215, USA\\
$^\$$ current affiliation: Department of Physics, Faculty of Science, University of Zagreb, Bijeni\u{c}ka c. 32, 10000 Zagreb, Croatia \\
$^\sim$ current affiliation: Quantum Artificial Intelligence Lab. (QuAIL), NASA Ames Research Center, Moffett Field, CA 94035, USA
KBR; 601 Jefferson St., Houston, TX 77002, USA}

\abstract{ A finite length ferromagnetic chain with opposite spin polarisation imposed at its two ends is one of the simplest frustrated spin models.  In the clean classical limit the domain wall inserted on account of the boundary conditions resides with equal probability on any one of the bonds, and the degeneracy is precisely equal to the number of bonds. If quantum mechanics is introduced via a transverse field, the domain wall will behave as a particle in a box, and prefer to be nearer the middle of the chain rather than the ends. A simple characteristic of a real quantum annealer is therefore which of these limits obtains in practice.  Here we have used the ferromagnetic chain with antiparallel boundary spins to test a real flux qubit quantum annealer and discover that contrary to both expectations, the domain walls found are non-uniformly distributed on account  of effective random longitudinal fields present  notwithstanding tuning carried out to zero out such fields when the couplings between qubits are nominally zero. We present a simple derivation of the form of the distribution function for the domain walls, and show also how the effect we have discovered can be used to determine the strength of the effective random fields (noise) characterising the annealer.  The noise measured in this fashion is smaller than what is seen during the single qubit tuning process, but  nonetheless qualitatively affects the outcome of the simulation performed by the annealer.}

\section*{Introduction}

The low energy states of natural systems can correspond to the solutions
of computationally difficult problems \cite{Barahona1982}. Experiments
suggest that these low energy states can be accessed and
measured by taking advantage of quantum mechanics, using a technique
known as quantum annealing \cite{Kadowaki1998,Brooke1999,Santoro2002}. Here, the "difficult" problem is converted to an equivalent problem of finding the ground state of an Ising Hamiltonian, and that classical ground state is approached via the introduction and subsequent removal of quantum fluctuations, typically imposed via transverse fields. It is suspected but by no means universally agreed that quantum annealing could provide an improvement over other methods for certain classes of interesting problems \cite{Farhi2001,Hogg2000}. 
To harness the power of quantum annealing, machines must be constructed
to faithfully implement the relevant transverse field Ising Hamiltonian (TFIM), and to do so represents a major challenge in quantum information science and engineering.  We refer to such machines as annealers. While the eventual outputs of annealers
usually take discrete binary values, the control parameters, which are the coupling constants in the TFIM, must be chosen
from a continuous set of values. An annealer should therefore be considered
an analog rather than a digital computer.  For a review of adiabatic quantum computing and quantum annealing see  \cite{Albash2018a} and for a forward looking perspective on the field see \cite{Hauke2020}.

Quantum annealing has attracted considerable experimental
attention recently \cite{Boixo(2013),Vinci(2014),Johnson2011,Boixo2013-1,Vinci2013,Harris2010,Santra2013,Coxson(2014)}, which is understandable given the wide variety
of applications, from traditional computer science problems \cite{Santra2013,Choi2010},
to more exotic uses such as aiding genetic algorithms to calculate
radar waveforms \cite{Coxson(2014)}, search engine ranking \cite{Garnerone2012},
graph isomorphism \cite{Vinci2013}, and portfolio optimization
\cite{Marzec}. In addition, sampling using a quantum annealer, which is effectively the topic of the current paper, is highly relevant to many machine learning and statistical inference tasks \cite{Chancellor(2016),Amin(2016),Benedetti(2016),Benedetti(2016)-1,Khoshaman(2019),Sadeghi(2019),Vinci(2019)}.

Precision of control parameters is a fundamental issue in analog computing, not present in its digital counterpart \cite{Bissel2004}. It is  
therefore important to ask what new complications these errors may
add. One could hope, for example, that small uncorrelated control errors
simply average out, leading to no noticeable effect
as long as they are below a threshold \cite{Young(2013)}. Long timescale noise should also be considered
a source of control error; this noise will be indistinguishable from
the TFIM being mis-specified by the device. We demonstrate here that the effects of control errors can be counter-intuitive, giving non-uniform distributions within a degenerate manifold
even for uncorrelated errors. We further argue that this effect captures
error-causing noise that would be missed if we try to measure the
errors with a different protocol. 

There is a growing literature on error correction in quantum annealing. Most of
the studies focus upon the effect of coupling to an external bath
rather than control errors \cite{Jordan(2006),Lidar2009,Lidar(2008),Young(2013)-1,Quiroz(2012),Ganti(2013)}.
The work in \cite{Young(2013),Pudenz(2014),Matsuura(2019)} does mention
techniques that can reduce the effect of control errors, at the cost
of some overhead, but cannot completely eliminate them. For the purposes of this study there are two kinds of relevant
noise processes, the dissipation which occurs on a time scale comparable or faster than the system dynamics, and slower noise
which appears fixed with respect to these timescales and acts as effective random field terms.
The role of the faster noise is to hasten relaxation to a thermal distribution,
while the slower noise (referred to as control errors) is what is directly measured in the experiments we report.

While the present study is dedicated to the consequences rather than the physical origins of the noise in flux qubit quantum annealers, we note other literature describing this noise as due to interactions of the qubit with magnetic defects in the chip substrate \cite{Kumar(2016)}. It typically takes a profile with a $\frac{1}{f}$ type frequency profile, meaning that the noise contains both high and low frequency components. The low frequency components can be treated as effectively static control errors, and are responsible for the effects that we report here.

We examine experimentally the effect of control errors on the annealer
constructed by D-Wave Systems, which mimics an Ising spin system.
Our experiment shows a non-uniform distribution within a ground state
manifold that can be explained by classical Boltzmann distributions
under the influence of field control errors, demonstrating that even
small errors affect the solution to strikingly simple problems. The fact that such a simple system subject to uniform noise can produce a non-uniform, but regular and predictable distribution, is an interesting mathematical result on its own, and to the best of our knowledge has not been previously reported in the literature. 

We find that even a domain wall in a one-dimensional system subject to
\emph{uncorrelated} field control errors yields a non-trivial U-shaped
domain wall distribution. An effective potential for the domain walls
is generated by combinatoric effects in the averaging over disorder
in the Hamiltonian. In this sense the phenomenon that we observe is due to an entropic potential. While the average domain-wall energy is the same at every site; there are more field configurations where the lowest energy spin configuration has the domain wall near the ends, the probability of observing it in such positions is higher. 
This is a finite size effect distinguishable from order-by-disorder which occurs in infinite systems, where the term was originally
used \cite{Villain1980} to describe entropic effects in the thermodynamic limit. We further demonstrate that this distribution can be
used to measure noise in the device and discuss the advantages over
the conventional method of examining single qubit auto-correlation.

The hardware that we use implements a transverse field Ising model
with a time-dependent Hamiltonian of the form,

\begin{equation}
H(t)=-A(t)\sum_{i=1}\sigma_{i}^{x}+B(t)\,H_{\mathrm{prob}}, \label{eq:H_anneal}
\end{equation}

where $H_{\mathrm{prob}}$ is a user-specified Ising Hamiltonian, which is diagonal in the $z$ basis, and $A(t)$,
$B(t)$ are the annealing schedule, the time dependences that control
the relative strength of each term.

\begin{figure}
\begin{centering}
\includegraphics[width=7cm]{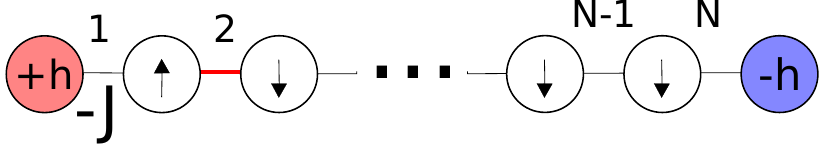}
\par\end{centering}

\caption{\label{fig:setup}Illustration of the Hamiltonian showing
one of the ground states assuming that $h>J$ and the qubits on the
end satisfy the applied fields. Numbers indicate domain-wall sites.}
\end{figure}

The specific problem that we choose to study is a ferromagnetic
chain with opposing fields at either end as shown in Fig.
\ref{fig:setup}. As long as $|h|>J>0$, $H_{\mathrm{prob}}$ will have an
$(N-1)$-fold degenerate ground state manifold consisting of all states
with a single domain wall; $\ket{\uparrow\downarrow\downarrow\ldots\downarrow}$
,$\ket{\uparrow\uparrow\downarrow\ldots\downarrow}$ ... $\ket{\uparrow\uparrow\ldots\uparrow\downarrow}$. This same system has been shown in \cite{Chancellor(2019),Abel(2021)} to be an effective method to encode discrete variables, and has been shown experimentally to improve performance in optimisation \cite{Chen(2021),Berwald(2021)}.
In our experiments, we use $h=2\:J$ with a Hamiltonian of the form
%
\begin{equation}
H_{\mathrm{prob}}=J\,\sum_{i=1}^{N}-\sigma_{i}^{z}\sigma_{i+1}^{z}+h\:(\sigma_{1}^{z}-\sigma_{N+1}^{z}).
\label{eq:particle-in-box_Ham}
\end{equation}

It is worth briefly noting that by keeping only the degenerate ground subspace of Eq.~\ref{eq:particle-in-box_Ham}, with the Hamiltonian in Eq.~\ref{eq:H_anneal}, we obtain a discretised particle-in-a-box Hamiltonian 
\begin{equation}
H_{\mathrm{deg}}=-A(t)\sum_{i=1}(a_ia^\dagger_{i+1}+a^\dagger_ia_{i+1}),
\label{eq:deg_subspace}
\end{equation}
where $a_i$ ($a^\dagger_i$) is an operator which annihilates (creates) a domain wall at location $i$. While in principle quantum effects within this manifold could be observed, we find that control errors dominate in our experiment.

We focus on control errors arising from stray magnetic fields from
free spins and dangling bonds within the materials that make up the
quantum processing unit (QPU). This could be considered equivalent to adding a term of the form,

\begin{equation}
H_{\mathrm{fields}}=\sum_{i}\,\zeta{}_{i}\,\sigma_{i}^{z},\label{eq:Hnoise}
\end{equation}

to the overall Hamiltonian, where $\zeta_{i}$ are uncorrelated and
Gaussianly distributed with a standard deviation $\sigma_{\zeta}\ll J$
and zero mean $\overline{\zeta_{i}}=0$. The overline indicates
an ensemble average. Random field terms such as those in Eq.~\ref{eq:Hnoise} appear naturally in implementations of the transverse field Ising model, including for example the dipole-coupled magnet LiHo$_{x}$Y$_{1-x}$F$_4$ \cite{Silevitch(2007)}. Because the coupling between the qubits and the substrate is likely to change with bias, $\zeta{}_{i}$ will generally be time-dependent, however as we show later the system remains in thermal equilibrium until very late in the anneal, so it may be treated as static for the purpose of these experiments.

 One can also consider coupler control errors
of the form $H=\sum_{i}\,\zeta_{i}^{(J)}\,\sigma_{i}^{z}\sigma_{i+1}^{z}$,
where $\zeta_{i}^{(J)}$ satisfy the same conditions as $\zeta_{i}$
with a standard deviation $\sigma_{J}\ll J$. This type of control
error produces an uncorrelated potential for the domain walls, and
therefore has no effect upon the shape of the mean thermal domain-wall distribution.
We demonstrate this in Sec. 2.1 of the supplemental material. It is
worth noting briefly that \cite{Johnson2011} describes a similar
experiment, but with the goal of demonstrating quantum tunneling, as had been done previously for a disordered magnet \cite{Brooke(2001)}. A qualitatively similar domain wall distribution to the one that
we see can be found in the supplemental material of \cite{Johnson2011}, although
there is no discussion of this finding and its possible origins. 

One concept that helps explain the behaviour of these systems is \emph{freeze time}, which is the time at which the dynamics of the QPU effectively stop and the spins are effectively fixed. Because the device appears to be reaching thermal equilibrium in these experiments, we can think of these experiments as measuring the ratio of the noise level to the device temperature at the freeze time. Since the susceptibility of flux qubits to external noise will, in general, be different at different points in the experiment, the freeze time is an important (but not directly measurable) parameter.

Beyond control parameter errors, there is the quantum mechanical effect of zero point fluctuations which could be frozen out in a sufficiently fast read-out  of the annealer; in the 1D problem considered by ourselves, the domain wall behaves as a "particle in a box" whose mass is regulated by the transverse field, implying a distribution of possible domain wall positions with a peak in the middle of the chain, unlike the flat distribution expected for the ideal classical limit. Therefore, quantum fluctuations can cause quantum annealers to sample -  in interesting ways -
ground states unfairly \cite{Matsuda(2009),Matsuda(2009)-1,Mandra(2017),Konz(2019),Kumar(2020)}, and such effects can sometimes be used in a beneficial way \cite{Zhang(2017),Chancellor(2020)}). However, we conclusively demonstrate that the effect observed here is classical in origin, both through numerical simulation and experiments using different runtimes.  The reason that this system is dominated by classical rather than quantum effects is that it equilibrates quickly through thermal barrier hopping, as has been independently observed for one-dimensional chains in quantum annealers in \cite{Izquierdo(2020)}.

\section*{Theoretical Analysis}

There is a vast literature on the one-dimensional Ising model dealing with issues ranging from  random fields in the classical limit \cite{Bruinsma(1983)}\, through  disordered couplers and transverse fields \cite{Fisher(1995)}, to twisted boundary conditions in the clean quantum limit (see e.g.  \cite{Campostrini(2015)}). Nonetheless, we could not locate a paper which specifically addresses the domain wall distribution for the one-dimensional Ising model with twisted boundary conditions and a longitudinal random field, and what is relevant for quantum simulators, the evolution of this distribution with a transverse field. While such domain-wall distributions  can be
easily obtained through numerical sampling as we describe below, we provide in this section
an analytical calculation in the classical limit, followed by some remarks on what could happen during the quantum annealing process.

Let us start by considering the energy
contribution from the field control errors in the case of a single
domain wall on the $n$th coupler in the chain, $E_{n}=\sum_{i=1}^{N}\textrm{sign}(n-i+0.5)\,\zeta_{i}.$
The difference in energy between two domain-wall positions is therefore
$E_{n}-E_{m}=2\,\sum_{i=m+1}^{n}\zeta_{i}$ where $n>m$.

Assuming that $\overline{\zeta_{i}}=0$ and $\bar{\zeta_{i}^{2}}=\bar{\zeta^{2}}$,
we note that $\overline{E_{n}-E_{m}}=2\,\sum_{i=m+1}^{n}\bar{\zeta}_{i}=0$,
but

\[
\overline{(E_{n}-E_{m})\,(E_{n}-E_{k})}
\]
\begin{equation}
=4\,\overline{\zeta^{2}}\,\min(|n-k|,|n-m|)\,\Theta\left[(n-k)(n-m)\right],\label{eq:not_same}
\end{equation}

where $\Theta$ is a Heaviside theta. Note that this formula explicitly
demonstrates that the domain wall energies are correlated, even for
uncorrelated fields. Also note that for a Gaussian distribution $\bar{\zeta^{2}}=\sigma_{\zeta}^2$.
The probability of finding a domain wall at site $n$ in a thermal ensemble averaged over noise is given by
$P_{n}=\overline{Z^{-1}e^{-\beta\,E_{n}}}=\overline{\left[1+\sum_{m\neq n}e^{-\beta\,(E_{m}-E_{n})}\right]^{-1}}$.

Let us now consider a high-temperature approximation to obtain an
analytical formula. By expanding this probability to second order
in $\beta=\frac{1}{k_B T}$ and applying Eq. \eqref{eq:not_same} we obtain 

{\footnotesize{}}%
{\footnotesize \par}

\begin{equation}
P_{n}\approx\tilde{P}+\beta^{2}\,\overline{\zeta{}^{2}}\,\frac{2}{N^{2}}\,\left(n-\frac{N+1}{2}\right)^{2}\label{eq:dw_dist_fields}
\end{equation}

where $\tilde{P}=\frac{1}{N}-\frac{\beta^{2}}{N^{3}}\overline{\zeta{}^{2}}(\frac{5}{4}N^{3}+N^{2}+\frac{1}{6}N+1)$.
This demonstrates that even small field control errors create a parabolic
(U-shaped) distribution of domain walls. Simple, uncorrelated errors
can have a strong effect on the equilibrium behavior
of a simple domain-wall system. Note that this calculation relies
upon the assumption that the system is in thermal equilibrium. We
justify this assumption numerically in Sec. 2.2 of the supplemental
material that accompanies this manuscript. We also demonstrate other
derivations at finite and zero temperature in Sec. 2.3 and 2.4 of
the supplemental material. The expansion used in Eq. \eqref{eq:dw_dist_fields}
is only guaranteed to be valid for temperatures much higher than the
maximum difference in domain-wall energies, $\beta\,\zeta\,\sqrt{N}\ll1$.
We therefore expect that this approximation will break down for long
chains, and experimentally demonstrate this breakdown in the paper. 
While the high temperature expansion provides valuable intuition, we perform all analysis by comparing to computer aided numerical calculations.

The phenomenon of Eq.~\ref{eq:dw_dist_fields} is interesting as an experimental tool, because it
provides a way of directly measuring the effect of the control errors
on the evolution of a non-trivial Hamiltonian (i.e. with non-zero interactions between qubits). Therefore,
we expect that the control errors measured in this way should
give a more accurate portrayal of the errors experienced in a real
computation than in single qubit methods where interactions between qubits have been set to zero.

We conclude - to motivate future research -  with considerations of what might happen during genuine (T=0) quantum annealing and quenches. Most noteworthy is that for the twisted boundary conditions represented by Eq.~\ref{eq:H_anneal} and Eq.~\ref{eq:particle-in-box_Ham} , we have a domain wall which can be thought of as a particle whose mass approaches zero and size (uncertainty in position) diverges as the quantum critical point where A=J is approached (see \cite{Brooke(2001)} and references therein). For a finite system, the wall is a particle in a box, which is more likely in its ground state to be found at the center of the box than at the edges. As we lower the transverse field below J, we expect the random fields to cause localization to occur as this is a one-dimensional system, i.e. in a chain of length L the domain wall should be localized as soon as its quantum mechanically defined size is smaller than L. What this means is that for a quantum annealer which is truly at zero temperature, the system can become trapped in a configuration which does not minimize energy. Furthermore, the random longitidinal fields would produce a distribution of wall positions (read out via projection of of individual spins onto the z axis) broader than expected for the clean limit.

\section*{Results}

\begin{figure}
\begin{centering}
\includegraphics[width=7cm]{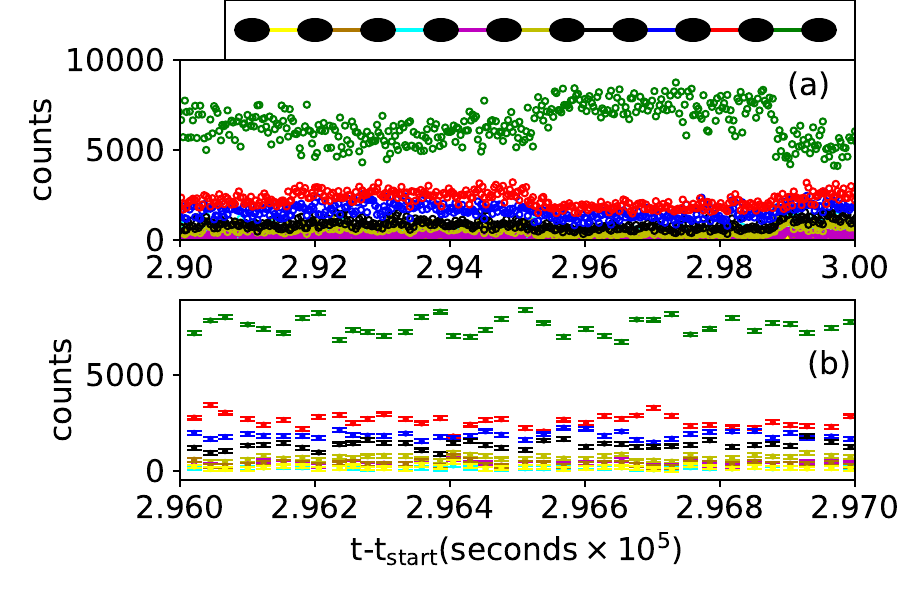}
\par\end{centering}

\caption{\label{fig:counts_vt}Experimentally observed domain-wall counts for a single
embedding and gauge choice versus sample time for a 10-qubit chain
(domain wall sites are color coded in cartoon). Time scales are different for (a) and (b). Error bars in the bottom frame are standard
error, and are suppressed in the top frame for clarity.
Dashed lines in the top figure represent times when the system updates
an internal self correction of biases. }
\end{figure}

Experiments were performed on a D-Wave Processor as described in Methods (and in more detail in the supplemental material), and we describe the key findings here. Firstly consider an individual instance of
the Hamiltonian shown in Fig. \ref{fig:setup} used in a quantum annealing protocol. Fig.~\ref{fig:counts_vt}
demonstrates  such an experiment, in particular annealing  on the same chain run repeatedly over time with no averaging either over different definitions of $0$ and $1$ on the qubits (gauge averaging), or over different physical chains. As with most other experiments reported here each anneal took $20 \mu s$. The distribution
of domain walls is non-uniform, as expected for local random fields even though we have tuned the qubit controls in an effort to eliminate such random fields.

We now check whether the simple classical considerations of the previous section can account for our experimental observations. To avoid effects due to local variations in qubits and couplers, we average over different chains and gauges.  Fig.~\ref{fig:counts_vt} also shows that the deviation between runs is much larger
than expected for identical samples drawn from the same distribution, which can be seen by comparing the actual spread of the points with the standard error depicted in the error bars in the lower frame.
This indicates that the control error has components that are faster
than the time between samples. Fast errors are more difficult
to detect, as well as to remove. For more discussion on this subject,
see Sec. 2.6 of the supplemental material.

The difference in domain-wall probabilities in Fig. \ref{fig:counts_vt}
is due to a combination of control errors, of the form given in Eq.
\ref{eq:Hnoise}, and coupler control errors. However, as we described in Sec. 2.1 of the supplemental material, measuring the average domain-wall distribution removes
the effect of coupler errors, allowing us to measure only the field
errors.

\begin{figure}
\begin{centering}
\includegraphics[width=7cm]{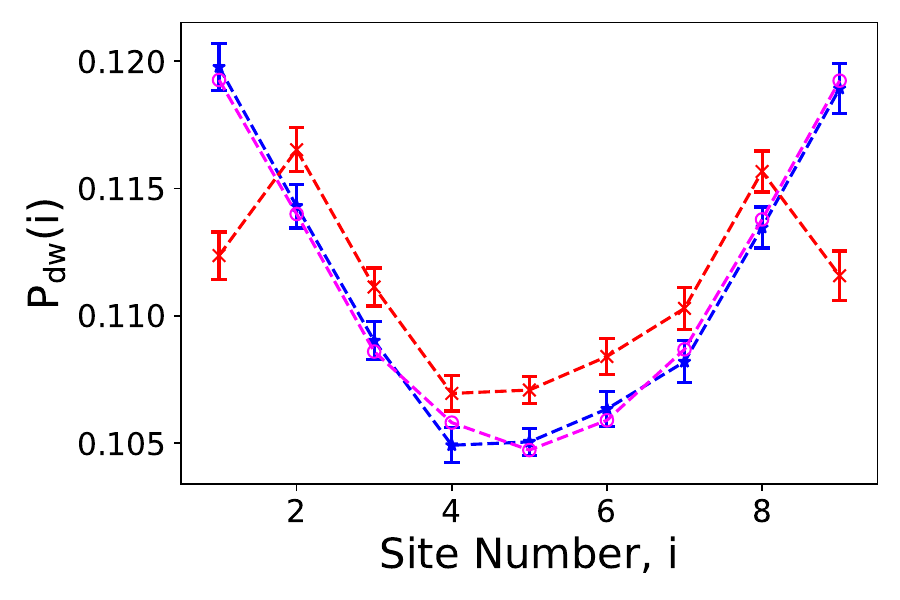}
\par\end{centering}

\caption{\label{fig:dw_dist} Domain-wall probability distributions for 10-qubit frustrated chain. Crosses are raw experimental data. Asterisks are
the same with a correction applied for background susceptibility.
Circles are numerically calculated data from sampling Boltzmann distributions
with field noise of the form Eq.~\eqref{eq:Hnoise} with $\frac{\sigma_{\zeta}}{T}=0.2363$. Lines joining points are a guide to the eye.}
\end{figure}

Fig. \ref{fig:dw_dist} displays the results from running the QPU
with the final Hamiltonian corresponding to the chain configuration
shown in Fig. \ref{fig:setup}, while averaging over many embedding
and gauge choices. An embedding corresponds to mapping a problem on a QPU such that every variable of the problem is represented by a subset of the qubits on the QPU. Note that chains can always be embedded in a one-to-one fashion, where every logical variable corresponds to one physical qubit; this is not true for more complicated graphs for which embedding is a more involved
process \cite{Fang(2020)}. Gauge choices arise due to an invariance of the target Hamiltonian under flips in the sign of a particular spin and the corresponding local field and couplings between it and other spins.  This averaging is explained in the Methods section
and Sec. 1.1 of the accompanying supplemental material. 

The experiment now
yields a U-shaped distribution, with the probability for the domain
wall to be located at the very end of the chain suppressed. The suppression
is predicted from well understood rf-SQUID background susceptibility
effects \cite{Harris(2010a)}, and can be removed by applying a simple linear correction; for more details see Sec. 1.2 of the supplemental material. Fig.~\ref{fig:long_chain} shows the behaviour of the experiments when performed on a longer chain, where the distribution deviates from parabolic due to the breakdown of the  assumptions underlying Eq.~\eqref{eq:dw_dist_fields}. For longer chains it is natural to ask whether Griffiths-McCoy-Wu singularities may be playing a role in the dynamics (such effects have recently been observed in two dimensional systems using quantum annealers \cite{Nishimura(2020)}). However, these would manifest themselves as unusual configurations such as multiple domain walls with regular spacings between them, rather than the distribution of single domain walls present in the ground state of a frustrated chain. In our  experiments we predominantly observed the single-domain-wall state indicating that these effects were not playing a crucial role.

\begin{figure}
\begin{centering}
\includegraphics[width=7cm]{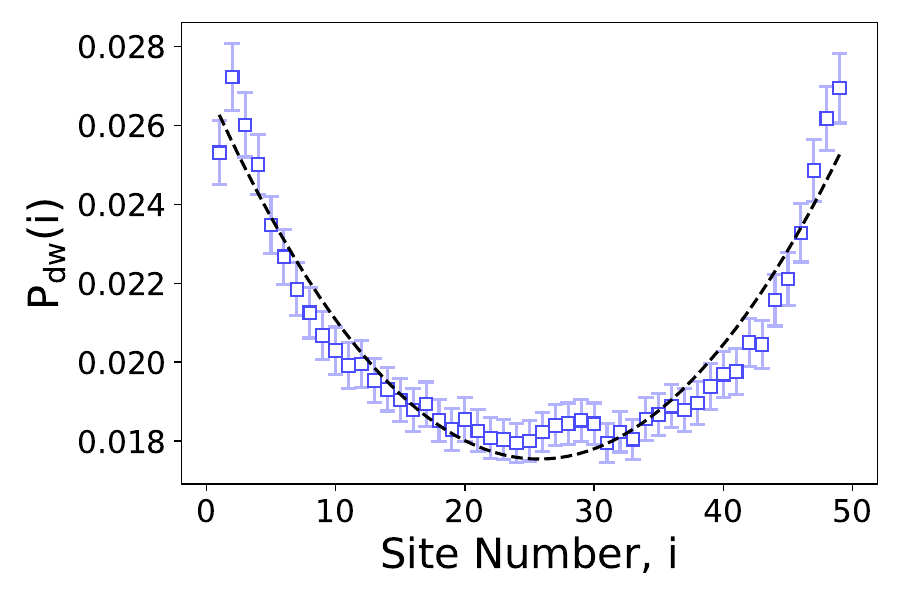}
\par\end{centering}

\caption{\label{fig:long_chain}Domain-wall distribution for a 50-qubit chain using the same experimental setup as Fig.~\ref{fig:dw_dist} (including background susceptibility corrections)
Dashed line is a parabolic fit. Error bars are standard error.}

\end{figure}

\begin{figure}
\begin{centering}
\includegraphics[width=7cm]{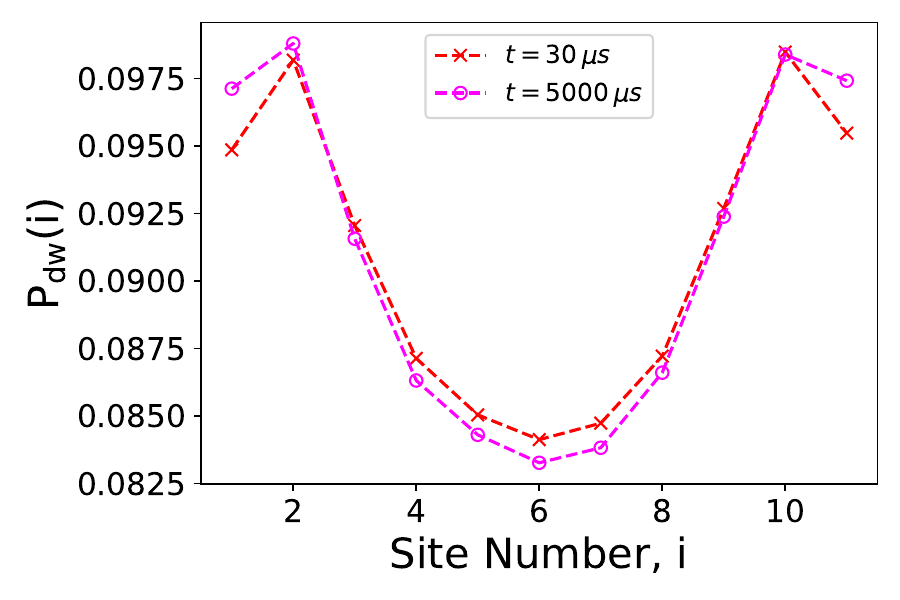}
\par\end{centering}

\caption{\label{fig:U_time} Domain-wall distribution for a 12 qubit chain with two different anneal times. Standard error errorbars are smaller than the depicted symbols. Background susceptibility corrections are not included.}

\end{figure}

Although the theoretical predictions provide a good fit to the data, it is important to establish whether we are really justified in treating the output as a classical thermal distribution, and whether any residual quantum effects remain. It has been shown that performing quantum simulations on spin chain systems using D-Wave annealers is difficult due to the fact that a very fast quench is required to capture the dynamics. In fact simulations performed on unfrustrated spin chains in \cite{Izquierdo(2020)} suggest that a quench would need to be of order $10^5$ times faster than currently available. As section 2.2 of the supplemental material shows, we find the same result. The simulations in \cite{Izquierdo(2020)} also suggest that the scaling of the experiments is not favourable with system size, indicating that it will actually be more difficult to observe quantum effects in longer chains than the short ones we have simulated. To further confirm that the system is very close to thermal equilibrium, we perform annealing at two very different annealing times, as depicted in Fig.~\ref{fig:U_time}. In this figure we see that making the run time orders of magnitude longer makes only a small difference in that the minimum of the distribution is slightly more pronounced at longer runtimes.

\begin{figure}
\begin{centering}
\includegraphics[width=7cm]{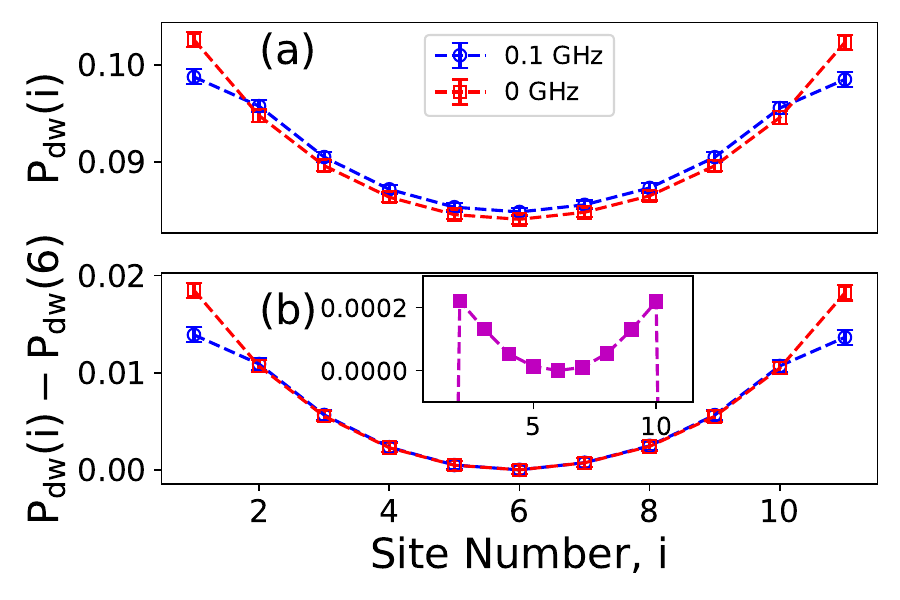}
\par\end{centering}

\caption{\label{fig:trans_field_eff} Top: Numerically calculated domain-wall distribution assuming thermal equilibrium in the presence or absence of a transverse field for experimentally realistic parameters . Bottom: same as top but with mid-point value subtracted to allow direct comparison, inset is difference between the two curves. Both data were taken using the same $10^5$ random noise realizations. Error bars are standard error, and this plot uses experimentally realistic parameters $\frac{\sigma_{\zeta}}{T}=\,0.24$ and  $T=15\, mK=0.31 \,GHz$}

\end{figure}

We next examine the effect of the weak transverse fields which are still present at the freeze time, and whether this can lead to interesting quantum effects. We assume that the freezing occurs when $A(t)=0.1\,GHz$, which is reasonable based
upon previous work \cite{Johnson2011}, and then compare the thermal distribution with or without the transverse field present, assuming a noise of $\frac{\sigma_{\zeta}}{T}=0.24$ which can be extracted by fitting our experimental data as described later. Assuming a temperature of $T=15\, mK=0.31\, GHz$, we find the distributions in Fig.~\ref{fig:trans_field_eff}. The transverse field has very little effect, aside from a slight suppression of terminal site probabilities. We therefore conclude that the dominant effects observed in these experiments are indeed classical, and perform our remaining analysis from the perspective of classical thermodynamics.

Returning to analysis of the 10-qubit chain, the experimental data match the numerical data obtained by Boltzmann
sampling over field noise of the type in Eq.~\eqref{eq:Hnoise} with
$\frac{\sigma_{\zeta}}{T}=0.24 \, (\sigma_{\zeta} \approx 0.074 \,GHz)$. This fitting was performed against
numerical sampling results rather than Eq.~\eqref{eq:dw_dist_fields}
to allow for higher-order corrections. Specifically, the fitting was performed by numerically sampling over disorder realizations and varying the parameter $\frac{\sigma_{\zeta}}{T}$ until the least squares error with respect to the experimental distribution was minimized. By contrast, a naive estimate
made by sampling uncoupled qubit polarization yields $\frac{\sigma_{\zeta}}{T}=0.13 \, (\sigma_{\zeta} \approx  0.040 \, GHz)$.

The result that a naive single decoupled qubit measure of local random fields is substantially below the random fields needed to generate the non-trivial "U" distribution of domain walls for coupled qubits is a key outcome of our study. One possible explanation is that each sample is averaged over many annealing
runs (each of which use the same annealing time as the domain wall experiments) and is therefore blind to any errors with a timescale less than
the time to collect all the samples, which is approximately 1 second. A more sophisticated
analysis based upon calculating autocorrelation via the Fourier transform (FT)
of the single-qubit results yields $\frac{\sigma_{\zeta}}{T}=0.35 \, (\sigma_{\zeta} \approx 0.11\, GHz)$
(see Sec. 1.3 of the supplemental material for details) and so together with the outcome of the polarization sampling technique, brackets the result obtained from the domain wall distribution for the Ising model with interacting qubits. As with the polarization calculation, the FT experiments are performed with the coupling turned off. This method is expected
to be sensitive to a wider bandwidth of errors than the naive measurement,
but should still be blind to any noise faster than the Nyquist interval,
which in this case is about $178\, \mathrm{\mu s}$. A summary of the different noise measurements can be found in table \ref{tab:measure_techniques}

\begin{table*}
\begin{centering}

\begin{tabular}{|c|c|c|c|c|}
\hline 
measurement technique & coupling on? & sensitive to minimum time scale & $\frac{\sigma_{\zeta}}{T}$ & $\sigma_{\zeta}$(GHz)\tabularnewline
\hline 
\hline 
Domain wall & Yes & N/A & 0.2363 & 0.074\tabularnewline
\hline 
Naive sampling & No & $1\,s$ & 0.13 & 0.040\tabularnewline
\hline 
Fourier transform & No & $178\,\mathrm{\mu s}$ & 0.35 & 0.11\tabularnewline
\hline 
\end{tabular}

\caption{\label{tab:measure_techniques} Summary of different noise measurement techniques and the results obtained from them.}

\par
\end{centering}
\end{table*}

Autocorrelation (FT), which has the fastest time scale, measures more error than the domain-wall technique or longer term sampling, which means that there are short-term fluctuations in uncoupled qubits which do not matter when we are dealing with the coupled qubits in the domain wall problem.

Methods based on measuring the U-shaped distribution can further be used to address questions about the source
of the control error. For example, we can measure the coupling dependence
of the field control error by fixing the gauge of the chain and averaging
over only embeddings. With a fixed gauge, we must use a different
embedding strategy to reduce correlations in the control errors caused
by embedding to qubits in the same unit cell, see Sec. 1.1 of the
supplemental material. As Fig. \ref{fig:comp-3gauge} demonstrates,
the depth of the U is different depending upon the gauge, which represents strong evidence that how the couplers are set influences the local random fields. This result is
consistent with measurements performed by others that indicate that ferromagnetic
couplers should couple more strongly to noise \cite{Lanting pc}.
Our method of measuring the coupler-dependent portion of the control error does not require operation
of the control lines outside of the preprogrammed annealing schedule
of the device, while the method employed by \cite{Lanting pc} does.

\begin{figure}
\centering{}\includegraphics[width=7cm]{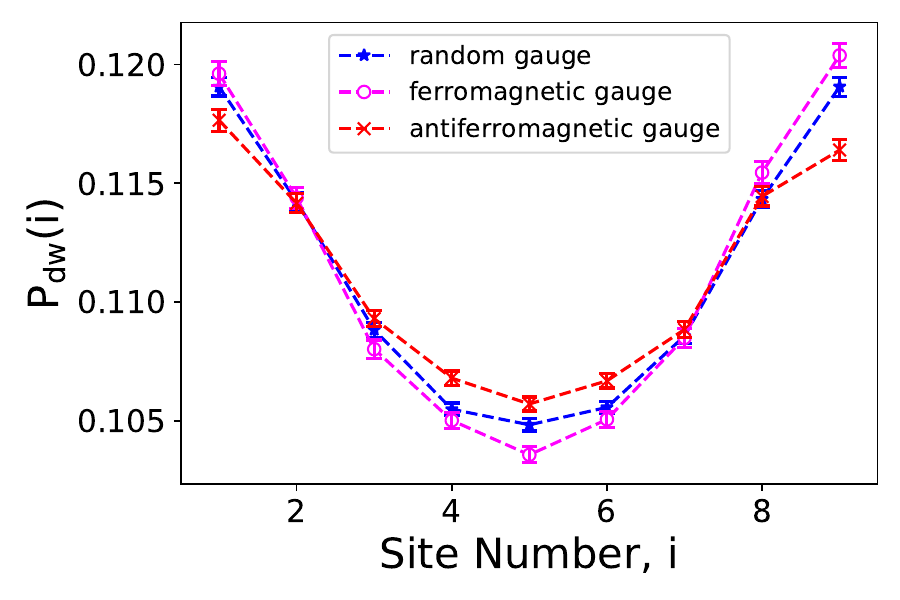}\caption{\label{fig:comp-3gauge} Domain-wall distribution with different gauge
choices. X represents data taken in the gauge which all couplers are antiferromagnetic.
Asterisks are averaged over random gauges. Circles are data for the gauge in which all couplers are ferromagnetic. All data in
this figure have been corrected for background susceptibility.}
\end{figure}

For the U-shaped distribution to be useful to measure field control errors requires the underlying assumption
that the errors are uncorrelated. Most types of correlation between
nearby qubits will be removed by the process of gauge averaging. However,
coupler-mediated errors from a shared coupler may depend on the state
(ferro or anti-ferro) of the coupler \cite{Lanting pc}, and therefore
may contain some correlations that survive gauge averaging. We suspect
that this part of the error should be relatively small because these
correlations will only come from one of the 5 or 6 couplers connected
to a given qubit, and only a fraction of the error from each coupler
is state dependent \cite{Lanting pc}.

\begin{figure}
\begin{centering}
\includegraphics[width=7cm]{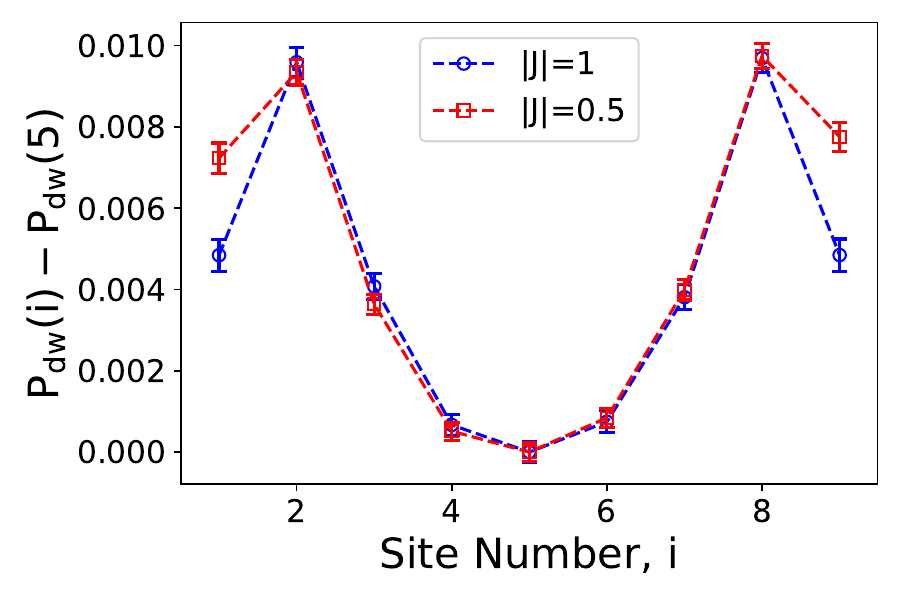}
\par\end{centering}

\caption{\label{fig:u_depth_scaleJ} Difference between domain wall probability
on site $i$ from the probability that domain wall is found on site
5 for two scales of the coupling. Note that background susceptibility corrections have not been performed.}

\end{figure}

We also have checked experimentally whether state-dependent errors
have a significant effect. Fig. \ref{fig:u_depth_scaleJ} demonstrates
that the depth of the distribution does not change within statistical
error when the strength of the coupling is reduced by a factor of 2.
If there were a strong component of the control error which depended
on the state of the couplers, we would expect a substantial difference
between the depth of these two distributions. This result is consistent with the mild dependence of the outcomes on gauge choice, and reinforces the concept that dynamical effects at the longest time and smallest energy scales are responsible for the discrepancies between the effective random fields seen for interacting and non-interacting qubits. 

\begin{figure}
\begin{centering}
\includegraphics[width=7cm]{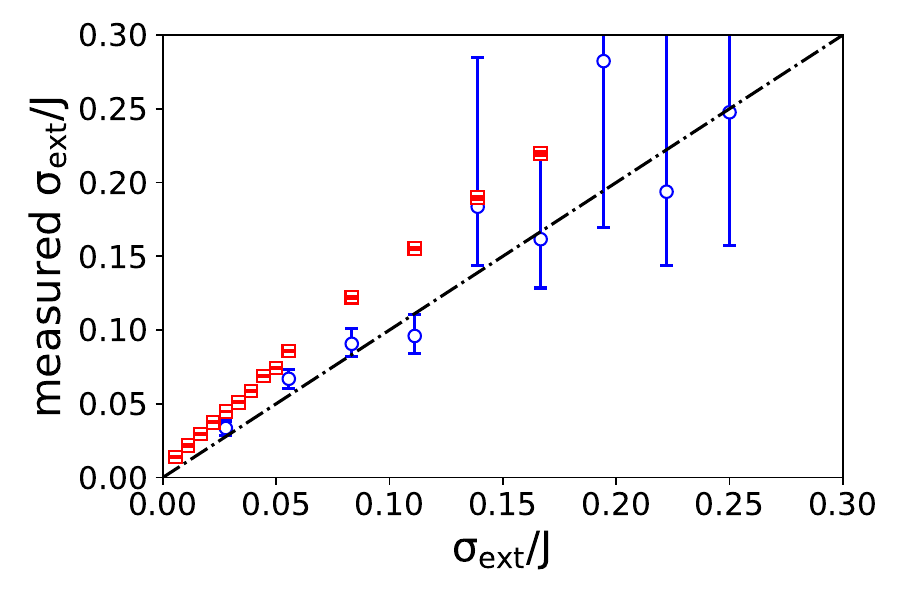}
\par\end{centering}

\caption{\label{fig:artificial_err}Plot of measured values of $\sigma_{ext}$, where the local random fields measured in the absence of the imposed local fields $\sigma_\mathrm{ext}$, are subtracted according to the quadrature formula in text, 
versus its actual value, assuming the system freezes when $A(t)=0.1\,GHz$. The effective random field strength we have used is the same as before, $\frac{\sigma_{\zeta}}{T}=0.24$, using at temperature of $15\,mK=0.31GHz$ and $J=1.80 GHz$, we obtain $\frac{\sigma_{\zeta}}{J}=0.041$.
Blue circles are from comparing the `depth' of the U with numerical
sampling. Red squares are from comparing autocorrelation with the
results of numerical sampling. Dot-dashed line is a guide to the eye
at equal measured and applied $\sigma_{ext}$. For this plot we define
the `depth' as the ratio of the probability to find the domain wall
in the middle site over the sum of all probabilities excluding terminal
sites. The terminal sites are excluded to avoid having to compensate
for the background susceptibility effect seen in Fig. \ref{fig:dw_dist}.}

\end{figure}

To validate the calibration of the random fields measured via the domain wall technique relative to the single qubit methods, we use the field controls
of the chip to insert Gaussian control errors with a width $\sigma_{ext}$.
Assuming that the original control error is not affected by the additional
error that we insert artificially, the two errors will be independent,
and the total error will be $\sigma_{tot}=\sqrt{\sigma_{\zeta}^{2}+\sigma_{ext}^{2}}$.
The strength of the inserted control errors, $\sigma_{ext}$ is dependent on $t$, in the annealing schedule specified by $A(t)$ and $B(t)$ and therefore the time at which the qubits become effectively
`frozen' .  It is important to note that $\sigma_{\zeta}$ can also
depend upon the freeze time, so different
freeze times with the couplings on versus off can potentially explain the deficit in the
errors measured by the chain method. For our analysis we assume that
the system freezes when $A(t)=0.1\,GHz$, which is reasonable based
upon previous work \cite{Johnson2011}. Fig. \ref{fig:artificial_err}
shows the measured value of $\sigma_{ext}$ versus the programmed value.
From this figure we see that within generous errors the domain wall distributions agree
with this model of the error, for which the results should appear on the diagonal indicated by the dashed blue line,  while local autocorrelation measures an excess
of error from the phenomenological result. 

\section*{Conclusions}

We consider the performance of a real annealer for one of the simplest illustrations of magnetic frustration, namely that of a magnetic domain wall in an Ising chain constrained to have opposite spin at either end. In the absence of random fields, the wall resides on any bond with equal probability. When quantum fluctuations are present, the wall behaves as a particle in a box, which is in its ground state will have maximum probability amplitude in the middle of the chain. On the other hand, when classical random fields dominate, the distribution function for the wall becomes ``U'' shaped, with a minimum at the middle of the chain. We demonstrate this result with a simple analytical calculation, and then proceed to observe that it is also the generic result for the D-Wave quantum annealer. Based on the fact that we see these effects even though local random fields are zeroed with couplings between qubits tuned to zero, new random fields may be induced when the couplers are turned on to implement interacting qubit Hamiltonians.  
Annealers are meant to solve optimization problems, e.g. in logistics or machine learning, with many degrees of freedom. The appearance of random fields as coupling terms are turned on after zeroing local random fields acting on individual qubits can produce the usual pathologies associated with random fields in statistical physics. Most notable among these are the pinning of "domain walls" \cite{Bruinsma(1984),Villain(1984)}.  
The possibility that the couplings introduce noise should not come as a surprise given that we are after all, dealing with an analog computer. It is also not unprecedented in realizations of quantum annealers: for example, local longitudinal random fields can be induced by transverse fields in magnetic systems (see \cite{Silevitch(2007)} and references therein). 
We can turn a problematic but  interesting effect (bug) into a benefit (feature) by using measurements of the ``U'' distribution to directly measure control errors for quantum annealers. 

This method has several advantages which makes it a useful tool for understanding control errors in quantum annealing. It allows measurements to be made when couplers are active, therefore providing a more realistic estimate of the effects of control errors when solving real problems. Furthermore, the tests require no special access beyond the ability to submit problems to the device, so will be applicable for cases where users with limited access to the controls want to characterize noise in the controls. The relative ease of performing the measurements coupled with the fact that the measurements are performed in a fundamentally different way than the standard single, decoupled qubit measurements will also make running the 1D domain wall problem a simple method for characterizing new devices. The errors detected for coupled qubits are important to characterize because it has been shown that if left unchecked, errors in the problem specification can have catastrophic effects on the result \cite{Pearson(2019)}. 

Our method runs with the standard
annealing protocol, requiring no privileged access to the control
lines, and measures the component of the noise which acts as control
error \emph{by construction}, with no frequency
cutoff that depends on the annealing time. The second point means that the method could
be used for arbitrarily long annealing times to observe deviations
from the user-specified fields during the annealing process. On the other hand, should a processor be claimed to be a quantum simulator with a sufficiently rapid quench and readout, the domain wall problem will yield a distribution with a maximum rather than minimum at the centre of the chain, thus providing a qualitative test as to whether the device is classical random-field or quantum fluctuation- dominated. The cross-over between random field and quantum fluctuation-dominated regimes has been observed for model magnets \cite{Schmidt(2014)}, and we look forward to seeing a demonstration for properly programmable quantum simulators such as arrays of Josephson junctions or ion traps. 

\section*{Methods}

The data in Fig.~\ref{fig:dw_dist} were taken on the USC Information
Sciences Institute Vesuvius 6 D-Wave QPU. Except where otherwise
stated, these data were averaged over gauges, as well as over ways
of embedding on the QPU. For more details about the embedding see
Sec. 1.2 of the supplemental material. Data in Fig. \ref{fig:counts_vt}
were taken using a QPU intermediate between the Vesuvius and Washington
QPU generations made available by D-Wave Systems Inc. Unless otherwise stated all data were
taken using an annealing time of $20\,\mathrm{\mu s}$. All individual data sets
are taken with 10,000 annealing runs.

\section*{Acknowledgments}

The authors would like to thank Trevor Lanting, Murray Thom, Anatoly
Smirnov, Stefan Zohren, and Jack Raymond for useful discussion. NC was supported
by Lockheed Martin corporation and EPSRC fellowship EP/S00114X/1 while completing this work. TD was
supported by EPSRC grant EP/K02163X/1. WV was supported by EPSRC grant
EP/K004506/1. AGG was supported by EPSRC grant EP/I004831/2. The authors
thank the University of Southern California for allowing access to
their D-Wave 2 quantum annealer as well as D-Wave Systems Inc. for
access to the intermediate-generation annealer. Collaboration with
USC is supported under EPSRC grant EP/K004506/1.

\section*{Author Contributions}

NC performed the experiments. MHA and NC performed the calculations
with useful assistance from AGG. MHA produced the simulations which
demonstrated that the system can be treated as equilibrated. NC wrote
the paper. WV performed early experiments which demonstrated the effect.
PJDC and TD performed simulations which informed the early directions
of the project. AGG recognized that the effect was related to order
by disorder. NC and GA designed the experiment with PAW providing a
particularly useful suggestion. All authors were involved in discussions
of the results.

\section*{Competing interests}

The authors have no competing interests to declare.

\section*{Data Availability}

The data are available from the authors upon request.

\end{document}